\documentclass[%
 reprint,
 amsmath,amssymb,
 aps,
]{revtex4-2}
\usepackage{amsmath}
\usepackage{amssymb}
\usepackage{physics}
\usepackage{graphicx}
\usepackage{dcolumn}
\usepackage{bm}
\usepackage{color}
\usepackage[normalem]{ulem}

\begin{document}
\preprint{AAPM/123-QED}

\title[Sample title]{Quantum beating and cyclic structures  in the phase-space dynamics of the Kramers-Henneberger atom}

\author{A. Tasnim Aynul$^1$, L. Cruz Rodriguez$^{1,2}$, C. Figueira de Morisson Faria$^{1,3}$ }
\affiliation{ $^1$Department of Physics and Astronomy, University College London, Gower Street, London WC1E 6BT, UK\\$^2$Department of Chemistry at the University of Warwick, Coventry
CV4 7AL, UK\\$^3$ICFO-Institut de Ciencies Fotoniques, The Barcelona Institute of Science and Technology, Av. Carl Friedrich Gauss 3, 08860 Castelldefels, Barcelona, Spain
}

\date{\today}

\begin{abstract}
We investigate the phase-space dynamics of the Kramers Henneberger (KH) atom solving the time-dependent Schr\"odinger equation for reduced-dimensionality models and using Wigner quasiprobability distributions. We find that, for the time-averaged KH potential, coherent superpositions of eigenstates perform a cyclic motion confined in momentum space, whose frequency is proportional to the energy difference between the two KH eigenstates. This cyclic motion is also present if the full time dependent dynamics are taken into consideration. However, there are time delays regarding the time-averaged potential, and some tail-shaped spilling of the quasiprobability flow towards higher momentum regions. These tails are signatures of ionization, indicating that, for the potential studied in this work, a small momentum spread is associated with stabilization. A comparison of the quasiprobability flow with classical phase-space constraints shows that, for the KH atom, the momentum must be bounded from above. This is a major difference from a molecule, for which the quasiprobability flow is confined in position space for small internuclear separation. Furthermore, we assess the stability of different propagation strategies and find that the most stable scenario for the full dynamics is obtained if the system is initially prepared in the KH ground state.
\end{abstract}

\keywords{Suggested keywords}
\maketitle
\section{\label{sec:intro}Introduction}
Light-induced potentials have been realized in many areas of physics, with a wide range of applications. For instance, optical lattices are instrumental in laser cooling and the production of Bose-Einstein condensates \cite{Grynberg2001,Greiner2008}, and have also been widely explored for quantum simulations \cite{Gross2017}. Furthermore, light has been used to shape potential energy surfaces and conical intersections and molecules \cite{Kubel2020}, and the concept of Floquet engineering, in which a periodic driving field is employed to create light-induced states and resonances in molecules, solids and nanostructures, has gained significant prominence \cite{Oka2017,Lucchini2022}.

Besides the light-induced potential barrier created by a strong, low-frequency field, through which an electron tunnels, which is a starting point in the physical mechanisms behind widely known phenomena in strong-field laser-matter interaction such as above-threshold ionization \cite{Becker2018,Becker2002Review,MilosReviewATI}, high-order harmonic generation \cite{Lewenstein1994} and nonsequential double ionization \cite{Faria2011,Becker2012}, perhaps the most traditional light-induced potential in attosecond science is associated with the Kramers Henneberger (KH) atom \cite{Henneberger1968}.
The KH potential is a light-induced dichotomous potential that has been extensively studied in the context of stabilization. In broad terms, stabilization means that, under specific intensity and frequency regimes, ionization will be suppressed for atoms in strong fields. The electronic wave packet is then trapped in this dichotomous potential, whose shape has often been linked to that of a molecule (see, e.g.,  \cite{Gavrila2002} for a review). Since the past decade, there has been a resurgence in studies of stabilization and the KH potential in connection with schemes to image the KH atom, using, for instance, IR and XUV pump-probe schemes \cite{Ivanov2022} and photoelectron spectroscopy employing velocity-map imaging \cite{Morales2011}. These studies were likely triggered by experiments that found that atoms may survive intensities of $10^{15} \mathrm{W}/\mathrm{cm}^2$ \cite{Eichmann2009}. Recently, theoretical studies have even reported that features encountered in molecules, such as quantum interference due to electron emission from spatially separated centers and resonance-enhanced ionization, are also present in the KH atom \cite{He2020}.

Thus, a legitimate question is how far the analogy between the KH atom and a molecule can be pushed, and how stable the light-induced potentials are.   To perform these studies, it is convenient to write the full Hamiltonian in an oscillatory frame of reference, which quivers with the electron in the presence of the laser field: the KH  frame. In the KH frame, the time dependence of the field is embedded in the binding potential. Thereby, an important indicator is how far the system’s dynamics are from that obtained for the time-averaged KH potential. This time-averaged potential is the zeroth term of the Fourier expansion of the time-dependent potential and has a dichotomous structure and eigenstates of well-defined shapes and parities.  A stable population in the KH eigenstates is evidence that the stabilization regime has been reached \cite{Gavrila1984,popov1999applicability}, as well as the electronic wave packet developing a double-peaked structure. Early studies using Floquet theory also indicate that the Floquet quasienergy states exhibit a structure resembling those of the time-averaged KH potential \cite{Sundaram1992}.  Different coherent superpositions of KH eigenstates can be obtained, for instance, by employing fields of different lengths \cite{Watson1996}.
 
Important tools for understanding a system’s dynamics are provided by the phase space (for a review see \cite{Chomet2021}). Besides providing information about the spatial localization of a time-dependent wave packet, phase-space tools also shed light on how its momentum distribution evolves in time, and whether its temporal evolution exhibits nonclassical features.  For instance, subtleties in the dynamics of strong-field ionization in atoms \cite{Czirjak2000,Zagoya2014,Dubois2018} and molecules \cite{He2008,Takemoto2010,Takemo2011,Chomet2019}, non-dipole time delays in tunneling ionization \cite{He2022} or high-order harmonic generation in inhomogeneous fields \cite{Zagoya2016}, and rescattering \cite{Kull2012,Czirjak2013,Baumann2015}, and how quantum effects facilitate enhanced ionization \cite{Chomet2019,Chomet2021} have been understood with the help of the phase space. Furthermore, Wigner quasiprobability distributions have been employed to characterize non-classical light \cite{Lewenstein2021}, and in laser-induced core dynamics \cite{Medisauskas2015,Busto2018} and attosecond interferometry \cite{Busto2022}.

Within strong-field atomic stabilization, quantum phase-space tools have been employed in early work, following the line of inquiry of whether stabilization was classical or quantum \cite{Sundaram1992,Bestle1993,Watson1995,Watson1996}, and to understand the signatures of wave packets driven by different field types \cite{Watson1996}. Tails in the Wigner quasi probability distributions were associated with ionization and a dichotomous structure was linked to the KH potential \cite{Watson1995,Watson1996}. Another area of investigation was the nonclassicality of time-dependent terms beyond the zeroth order, associating multiphoton transitions with the corrections to the KH potential \cite{Bestle1993}, time-averaged potentials and scarring in the phase-space distributions \cite{Sundaram1992}.  
In addition to those, there are also studies using the classical phase space, employing periodic orbit theory and Poincar\'e sections \cite{Norman2015,Floriani2024}. These studies show that periodic orbits form if the electron drift momentum vanishes, and assess the role of resonances in stabilization. Vanishing momentum transfer has also been identified as a condition for stabilization using functional analysis methods in a quantum framework \cite{Fring1996,Faria1998b,Faria1998c}. 
Nonetheless, comparatively little attention has been paid to characterizing the temporal evolution of the KH eigenstates, or coherent superpositions thereof, in phase space. How do these coherent superpositions evolve in time, for a time-averaged or full potential in the KH frame, and what is the physics behind it? To what extent can this evolution be compared with that observed for molecules? Thereby, a key issue is that, if there is population trapping in a molecule, the quasiprobability flow undergoes a cyclic motion, in which direct transfer from one center to the other is observed \cite{He2008,Takemoto2010,Takemo2011,Chomet2019}. This flow is non-classical and associated with quantum interference, which facilitates this transfer. Another pre-requisite for this cyclic motion is spatial confinement, which in phase space is characterized by nested separatrices \cite{Chomet2019}.  The frequency of this cyclic motion is determined by the molecular parameters \cite{Kufel2020}, and can be controlled by an appropriate choice of the driving field parameters  \cite{Chomet2022}. 

In the present article, we investigate the phase-space dynamics in the KH frame and show that the same cyclic behavior is present for the KH atom if the stabilization regime has been reached. However, a key difference from a molecule is the lack of spatial confinement. A wave packet constructed as a coherent superposition of KH eigenstates will exhibit a cyclic motion between the two wells of the KH time-averaged dichotomous potential. A phase-space analysis employing Wigner quasiprobability distributions in the KH frame for a short-range potential shows that, while the position spread of the quasiprobability flow is large, a small spread around vanishing momentum is observed in the regime for which stabilization occurs. Stabilization breaks down once this momentum starts to increase, and the full time-dependent dynamics lead to a delay in this cyclic motion. A thorough understanding of these dynamics is important for creating stable light-induced coherent superpositions of states. 

This article is organized as follows. In Sec.~\ref{sec:backgd}, we provide the necessary theoretical background to understand our results and an outline of how our model is implemented. Subsequently, in Sec.~\ref{sec:proof}, we perform a series of tests in the key observables to ensure stabilization occurs. In Sec.~\ref{sec:phase-space} we proceed to a phase-space analysis, first characterizing the dynamics of KH eigenstates and then assessing the phase-space behavior associated with coherent superpositions of KH eigenstates. Finally, in Sec.~\ref{sec:conclusions} we state our conclusions. 

\section{\label{sec:backgd}Background}

\subsection{General expressions}
\label{sec:tdse}
In this section, we present the theoretical formulation to study an atom under the influence of a strong external field. We consider a one-dimensional model atom with a single active electron in an intense laser field, whose evolution is given by the time-dependent Schr\"odinger equation (TDSE)
\begin{equation}\label{eq:tdse}
    i\dfrac{\partial \psi_L(x,t)}{\partial t}=H_L(t)\psi_L(x,t),
\end{equation}
where $\psi_L(x,t)$ is the time-dependent wave function, and the time-dependent Hamiltonian in the length gauge and the dipole approximation reads
\begin{equation}\label{eq:hamiltonian}
H_L(t)=\dfrac{\hat p^2}{2}+V(\hat{x})-\hat{x}\varepsilon(t).
\end{equation}
 In the above equation, $V(\hat{x})$ is the binding potential, $\varepsilon(t)$ is the time-dependent electric field and the hats indicate operators. We use atomic units throughout, for which $m=e=\hbar=1$ and $\hat p=-i\partial/\partial x$ is the momentum operator in the position representation.

To model the laser-atom interaction for high-frequency and high-intensity fields, it is useful to employ the Kramers-Henneberger (KH) gauge. We obtain the Hamiltonian and the wave function in the KH gauge by applying the unitary transformation \cite{Henneberger1968,Faria1999,Richter2016}
\begin{equation}\label{eq:unitary}
    \mathcal{T}_{L\rightarrow KH} = \exp[i\int^t_0 A^2(\tau)d\tau]\exp[i\alpha(t)\hat{p}_x]\exp[-iA(t)\hat{x}],
\end{equation}
where $A(t)=-\int_0^t \varepsilon(\tau)d\tau$ is the vector potential, and $\alpha(t) = \int_0^t A(\tau)d\tau$ is the electron classical displacement in the electromagnetic field. One should note that this gauge transformation also changes the frame of reference. Thus, it takes the system from the length gauge, which is conventionally associated with the lab frame, to the Kramers-Henneberger reference frame (KH frame), whose origin oscillates as an electron moving freely under the action of the electromagnetic field. Therefore, throughout this work, we refer to the `laboratory frame' when employing the Hamiltonian given by Eq.~\eqref{eq:tdse} in the length gauge, and to the `KH frame' when using the KH gauge Hamiltonian and its approximations, unless a more rigorous classification is needed. 

The time-dependent Schr\"odinger equation in the KH frame takes the form
\begin{equation}\label{eq:HKH}
i\dfrac{\partial \psi_{KH}(x_{KH},t)}{\partial t} = \biggl(\dfrac{\hat p^2}{2}+V(x_{KH}+\alpha(t))\biggr)\psi_{KH}(x_{KH},t),
\end{equation}
\noindent where the term in brackets denotes the KH-frame Hamiltonian, $\psi_{KH}(x_{KH},t)$ give the KH-frame wave function and we have used $\hat {p}=\hat {p}_{{KH}}$ and $x_{KH}\ =x+\alpha(t)$. Note that we have dropped the hat in $x$ after applying the transformations, as our framework is the position representation.  

The above equations describe the electron dynamics in the KH reference frame, where the potential is now shifted by $\alpha (t)$ which defines the quiver motion of the electron. In this frame, the potential is periodic and can be expanded in a Fourier series \cite{Richter2016}
\begin{multline}\label{eq:fourier}
    V(x_{KH}+\alpha(t))=\sum_{n=-\infty}^{+\infty} V_n(x_{KH};\alpha_0)e^{in\omega t}\\=V_0(x_{KH};\alpha_0)+\sum_{n\ne 0}V_n(x_{KH};\alpha_0)e^{in\omega t}.
\end{multline}
The $n^{th}$ harmonic of the Fourier expansion is given by,
\begin{equation}\label{eq:nfourierterm}
    V_n(x_{KH};\alpha_0)=\dfrac{1}{T}\int_0^T V(x_{KH}+\alpha(t))dt e^{-in\omega t},
\end{equation}
\noindent where $T$ and $\omega=2\pi/T$ are the period and frequency of the field, respectively, and $\alpha_0=\varepsilon_0/w^2$ is the amplitude of the electron's quiver motion in a field of strength $\varepsilon_0$. 
The KH approximation assumes that the zeroth-order term
\begin{equation}\label{eq:fourierterm}
    V_0(x_{KH};\alpha_0)=\dfrac{1}{T}\int_0^T V(x_{KH}+\alpha(t))dt
\end{equation}
dominates the dynamics and the higher-order terms of the expansion can be assumed as a small perturbation. Under this approximation, the KH Hamiltonian takes the form
\begin{equation}\label{eq:Kh average}
    i\dfrac{\partial \psi_{KH}(x_{KH},t)}{\partial t} = \biggl(\dfrac{\hat p^2}{2}+V_0(x_{KH};\alpha_0\biggr)\psi_{KH}(x_{KH},t),
\end{equation}
and Eq.~\eqref{eq:fourierterm} defines the KH time-averaged potential.
This approximation renders the Hamiltonian time independent, with eigenstates $\phi^{KH}_n(x)$ \cite{Gavrila1984,popov1999applicability}.
The KH eigenstates are obtained by solving the time-independent Schr\"odinger equation
\begin{equation}\label{eq:KHeigenstates}
      \hspace*{-0.3cm}\biggl(\dfrac{\hat p^2}{2}+V_0(x_{KH};\alpha_0)\biggr)\phi_n^{KH}(x_{KH})=E^{KH}_n\phi_n^{KH}(x_{KH}),
\end{equation}
where $E^{KH}_n$ is a generic eigenenergy associated with the $n^{th}$ KH eigenstate $\phi^{KH}_n(x)$.
If the KH approximation is valid, the basis set formed by the eigenstates of the KH Hamiltonian \eqref{eq:KHeigenstates} is the most suitable to describe the electron dynamics. In the next section, we will obtain the eigenstates of the KH time-averaged potential for the specific potential and field parameters employed in our model.

 In this work, we solve both the full TDSE using Eq.~\eqref{eq:tdse} written in the laboratory frame, and Eq~\eqref{eq:Kh average} using the KH time-averaged potential explicitly given in the KH frame. To perform meaningful comparisons, the wave functions obtained in the lab frame after solving Eq.~\eqref{eq:tdse} are transformed to the KH frame by applying the unitary transformation in Eq.~\eqref{eq:unitary}. The wave function in the KH frame evaluated at the grid points $x_i$ in the lab frame has the form
\begin{multline}\label{eq:psiKH}
    \psi_{KH}(x_i,t) = \exp \biggl(\frac{i}{2} \int_0^{t}A^2(\tau)d\tau-iA(t)(x_i+\alpha(t))\biggr)\\
    \psi_L(x_i+\alpha(t),t),
\end{multline}
\noindent where the operators have been applied accordingly. From the numerical solution of equation \eqref{eq:tdse} we obtain the wave function $\psi_L(x_i,t)$, hence a spatial translation is performed to obtain $\psi_L(x_i+\alpha(t),t)$. The exponentials in Eq.~\eqref{eq:psiKH} require the evaluation of the integral of the electric field and the vector potential. We use definite integrals to ensure the classical displacement and momentum transfer at the beginning and the end of the pulse are zero. This is necessary to guarantee that no artifacts are introduced in the calculation and that stabilization may occur \cite{Fring1996,Faria1998b,Faria1998c}. 

To show the validity of the KH approximation under the parameters used in this paper we compute observables such as the time-dependent population of the initial atomic bound state $\phi_b(x)$ \cite{popov1999applicability}
\begin{equation}\label{eq:population_atomic}
    \mathcal{P}_{b}(t) = \biggl|\int \psi_L(x,t)\phi^*_b(x) dx\biggr|^2,
\end{equation}
and the time-dependent population of the KH eigenstates, given by 
\begin{equation}\label{eq:population_KH}
   \mathcal{P}_{KH}(t) = \biggl|\int \psi_{KH}(x,t)\phi^{*KH}_i(x) dx\biggr|^2,
\end{equation}
where the wave function $\psi_{KH}(x,t)$ is obtained by applying the transformation \eqref{eq:psiKH} after solving the TDSE given by   Eq.~\eqref{eq:tdse}. The states $\phi^{KH}_i(x)$ are the KH eigenstates evaluated from Eq.~\eqref{eq:KHeigenstates}.

It is important to notice how these observables depend on the reference frame used for the calculations due to the phases included in the wave function. Observables like the probability density in the KH frame take the form
\begin{equation}
    \biggl| \psi_{KH}(x,t)\biggr|^2=\biggl| \psi_{L}(x+\alpha(t),t)\biggr|^2,
\end{equation}
which means that, in the KH frame, the probability density will oscillate within the turning points of the KH time-averaged potential.

\subsection{Phase-space tools}
Here, we outline the phase-space tools used in this work: Wigner quasiprobability distributions and classical phase portraits. In our studies, it suffices to employ  Wigner quasiprobability distributions in the KH frame, due to the Galilei invariance of the KH transformation \cite{Watson1996}. However, for completeness, we state their expressions in both KH and length gauges. Although the Wigner quasiprobability flow is quantum, classical phase portraits provide insights into position and momentum constraints and indicate whether they are followed by the quasiprobability flow.
\subsubsection{Wigner quasiprobability distribution}
The Wigner quasiprobability distribution for a pure state $\ket{\psi_{G}}$ \cite{schleich2011quantum} reads
\begin{multline}\label{eq:wigner}
 W_{G}(x, p, t)=\int_{-\infty}^{\infty} \mathrm{d} \xi \bra{\psi_{G}}\ket{x+\xi/2}\bra{x+\xi/2} \ket{p}\\
 \bra{p}\ket{x-\xi/2} \bra{x-\xi/2}\ket{\psi_{G}},
\end{multline}
where the subscript $G=L$ or $KH$ indicates the gauge of the space-state vector and the Wigner function. The Wigner function is real and normalized, 
and its marginals correspond to physical probability distributions for each conjugate variable, respectively. This implies that the marginals are gauge-independent, although the Wigner function is not. The function $W_{G}(x,p,t)$ may assume negative values, and this can indicate nonclassicality. In terms of wave functions, Eq.~\eqref{eq:wigner} reads
\begin{equation}\label{eq:wigner2}
W_G(x, p, t)=\frac{1}{2\pi} \int_{-\infty}^{\infty} \mathrm{d} \xi \psi_G^{*}(x+\xi/2, t) \psi_G(x-\xi/2, t) \mathrm{e}^{\mathrm{i} p \xi}.
\end{equation}

The length-gauge Wigner function $W_L(x, p, t)$ can be written in terms of $\psi_{KH}(x,t)$ using the inverse of the transformation given by Eq.~\eqref{eq:unitary}, and reads
    \begin{multline}\label{eq:wignerL}
W_L(x, p, t)=\frac{1}{2\pi} \int_{-\infty}^{\infty} \mathrm{d} \xi \psi_{KH}^{*}(x+\xi/2-\alpha(t), t) \\\psi_{KH}(x-\xi/2-\alpha(t), t) \mathrm{e}^{\mathrm{i} [p-A(t)] \xi}.
\end{multline}
Similarly, the Wigner function $W_{KH}(x, p, t)$ in the KH gauge reads
 \begin{multline}\label{eq:wignerKH}
W_{KH}(x, p, t)=\frac{1}{2\pi} \int_{-\infty}^{\infty} \mathrm{d} \xi \psi_{L}^{*}(x+\xi/2+\alpha(t), t) \\
\psi_{L}(x-\xi/2+\alpha(t), t) \mathrm{e}^{\mathrm{i} [p+A(t)] \xi},
\end{multline}
if expressed in terms of the length-gauge wave function. Eq.~\eqref{eq:wignerKH} is extensively used in this article when assessing the full time-dependent dynamics.
\subsubsection{Phase portraits}
\label{sec:classical}
Phase portraits are useful to identify different phase-space behaviors, for instance, bound or continuum dynamics, or confinement. They are used in the framework of classical dynamical systems.  They are defined by solving Hamilton's equations
\begin{eqnarray}
    \dot{x}&=&p  \notag\\
    \dot{p}&=&-dV/dx.
\end{eqnarray}
A fixed point is obtained for $\dot{x}=\dot{p}=0$ and, for Hamiltonian systems, are either centers or saddles. If trajectories propagate along closed orbits, they occupy a bound region of the phase space. For overall definitions see Ref.~\cite{Arrowsmith1992}.

For time-independent Hamiltonian systems, it is convenient to define equienergy curves, given by 
\begin{equation}
    H(x,p)=E,
    \label{eq:equienergy}
\end{equation}
where $E$ is the energy of the system and $  H(x,p)$ the classical Hamiltonian. If an equienergy curve delimits phase-space regions of different dynamics, it is called a separatrix. Separatrices set boundaries for, for instance, bound and continuum dynamics, or oscillatory and rotational modes. Sometimes one can use equienergy curves approximately for time-dependent systems, but this approximation may not be reliable if the system changes too quickly (see \cite{Zagoya2014,Chomet2019,Chomet2022} for examples in low-frequency fields). 

\subsection{Model and numerical implementation}
For the present study, we choose the short-range potential 
\begin{equation}\label{eq:potential}
    V(x)=V_0\dfrac{\exp(-(x^2+16)^{1/2})}{(x^2+6.27^2)^{1/2}},
\end{equation}
\noindent where $V_0=-24.856 \rm \ a.u$. For these particular parameters, there is only one bound state with energy $E_g=-0.0276 \rm \ a.u$ ($-0.75 \rm \ eV$) \cite{grobe1993wavepacket,popov1999applicability}.
The ground-state energy and wave function $\phi_b(x)$ are obtained by imaginary time propagation neglecting the electric field in Eq.~\eqref{eq:tdse}. 

The electric field has the form 
\begin{equation}\label{eq:efield}
    E(t)=\varepsilon_0f(t)\sin(\omega t),
\end{equation}
\noindent where $f(t)$ is a trapezoidal envelope with two ramps on/off periods $T_1=2T$,
\begin{equation}\label{eq:envelope}
     f(t)=
     \begin{cases}
     t/2T_1       &  0\le t\le T_1\\
     1           & T_1\le t\le 22 T,\\
     -1/2T(t-T_f) &  22T<t<T_f
     \end{cases}
\end{equation}
\noindent where $T_f=24T$ is the total duration of the pulse. The field intensity is $I=5.7\times 10^{13} \rm \ W/cm^2$ and the period is $T=100\rm \ a.u$, which corresponds to a wavelength around $727 \rm \ nm$ and a frequency $\omega=0.0628 \rm \ a.u$.

\begin{figure}[hbt!]
    \centering
    \includegraphics[width=1\linewidth]{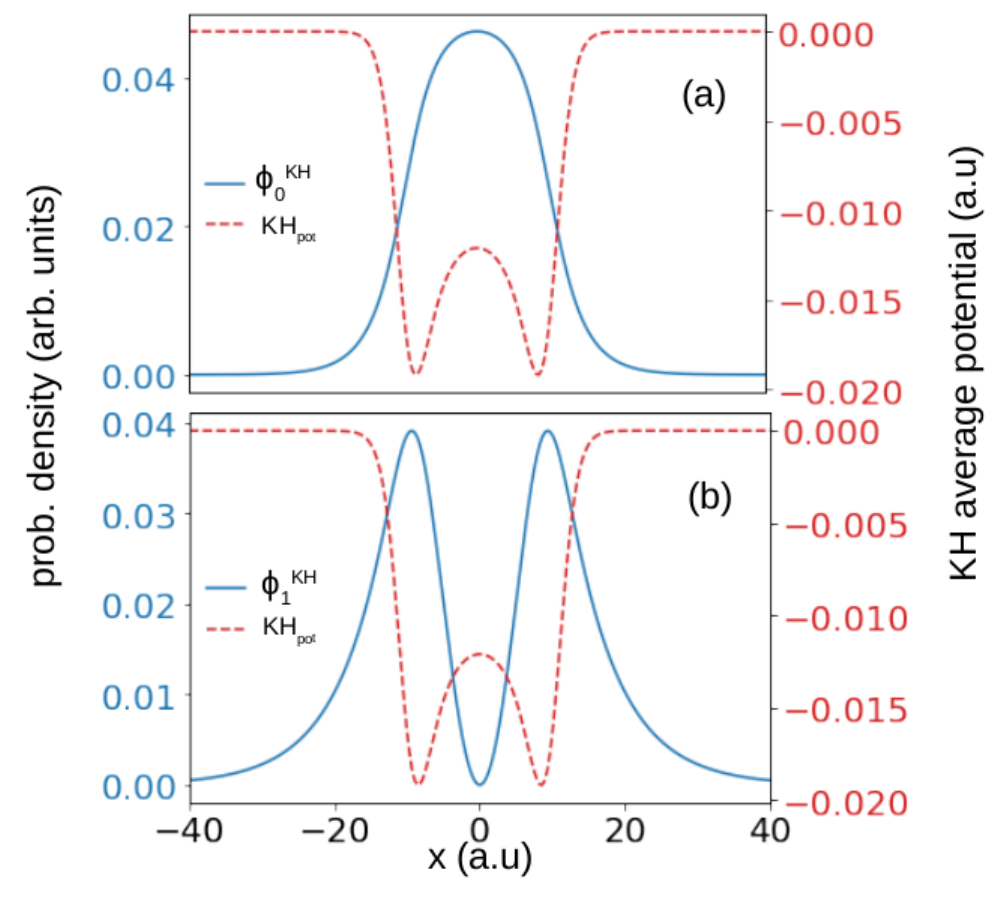}
    \caption{Probability densities associated with the KH eigenstates, together with the time-averaged KH potential used in this article. Panel a) displays the probability density $|\phi^{KH}_0(x)|^2$ associated with the ground eigenstate of the KH time-averaged potential, and panel (b) presents the probability density $|\phi^{KH}_{1}(x)|^2$ related to the excited KH eigenstate. In both panels the red dashed line represents the KH time-averaged potential obtained by numerical integration of Eq.~\eqref{eq:Kh average}. The eigenenergies of the ground and excited KH eigenstates are $E^{KH}_0=-0.01098 \rm \ a.u $ and $E^{KH}_1=-0.00282 \rm \ a.u $, respctively, and the maximum energy of the central KH potential barrier is $E_{\mathrm{sep}}=-0.0115 \rm \ a.u $.  }
    \label{fig:1}
\end{figure}

The KH time-averaged potential, whose general expression was given in Eq.~\eqref{eq:fourierterm}, is computed by numerical integration for the specific atomic potential in Eq.~\eqref{eq:potential}. This potential is shown in both panels in Fig.~\ref{fig:1} (red dashed line). As expected, the KH time-averaged potential is dichotomous, with wells centered at $x+\pm \alpha_0$, where $\alpha_0=\varepsilon_0/\omega^2$ is the amplitude of the electron quiver motion. The eigenfunctions and eigenenergies of the KH time-averaged potential are obtained by numerically solving the time-independent Schr\"odinger equation Eq.~\eqref{eq:KHeigenstates} using a discretization method as described in Ref.~ \cite{javanainen1988numerical} imposing the wave function vanishes at the edges of the grid. For the chosen potential and field parameters there are only two eigenstates of the KH time-averaged potential. This agrees with the results presented in Ref.~\cite{vivirito1995adiabatic} showing how short-range potentials support few eigenstates of the KH potential.

In Fig.~\ref{fig:1}, panels (a) and (b), we plot the probability density of the ground $|\phi^{KH}_0(x)|^2$ and first excited KH eigenstates $|\phi^{KH}_{1}(x)|^2$ (blue solid line), respectively. The ground state wave function is even and has a maximum located at $x=0$, while the first excited state wave function is odd and has extrema located at the wells of the time-averaged KH potential. 
The associated eigenenergies are $E^{KH}_0=-0.01098 \rm \ a.u $ and $E^{KH}_1=-0.00282 \rm \ a.u $, and are above the maximum energy of the central KH potential barrier, which is $E_{\mathrm{sep}}=-0.0115 \rm \ a.u $.  The KH potential minima are located at approximately $\pm\alpha_0=\pm10.23$ a.u. The parities of these eigenstates and the shapes associated with the corresponding probability densities resemble those observed in a diatomic molecule. The ground KH eigenstate has even parity and the associated probability density is mainly localized at the midpoint between both wells, while the excited KH eigenstate has odd parity, with the corresponding probability density peaked at the wells. 

We model the full quantum dynamics by solving the TDSE given in Eq.~\eqref{eq:tdse} in the lab frame. The equations are solved using the split operator method in a grid from $x\in [-1500:1500] \rm \ a.u$ . An absorbing potential is applied in the region outside $x\in [-600:600] \rm \ a.u$ to avoid reflections at the edges of the grid.
Additionally, we will perform computations for the KH time-averaged Hamiltonian \eqref{eq:Kh average}, which is intrinsically in the KH frame. The same grid and absorbing boundary conditions as for the full TDSE solution, in which the Hamiltonian is not approximated, are used in this calculation.  Our region of interest spans from $-60 \hspace{0.1cm}\mathrm{a.u.}\leq x\leq 60 \hspace*{0.1cm}\mathrm{a.u.}$, which is around six times the electron excursion amplitude.

We also use the time-averaged KH potential to compute classical equienergy curves in phase space, determined by
\begin{equation}
    E=\frac{p^2}{2}+  V_0(x_{KH};\alpha_0).
\end{equation}

The Wigner quasi-probability distribution is computed using the propagated time-dependent wave function as input in Eq.~\eqref{eq:wigner2}. For the time-averaged KH potential, the wave function is extracted directly from  Eq.~\eqref{eq:Kh average}, while, for the full potential, it is first computed in the length gauge using Eq.~\eqref{eq:tdse} and then transformed to the KH frame using Eq.~\eqref{eq:psiKH}. The Wigner function is computed for the spatial range of interest, namely $-60 \hspace{0.1cm}\mathrm{a.u.}\leq x\leq 60 \hspace*{0.1cm}\mathrm{a.u.}$

\section{Proof of concept}
\label{sec:proof}
\begin{figure}[hbt!]
\hspace*{-1cm}
    \includegraphics[width=1\linewidth]{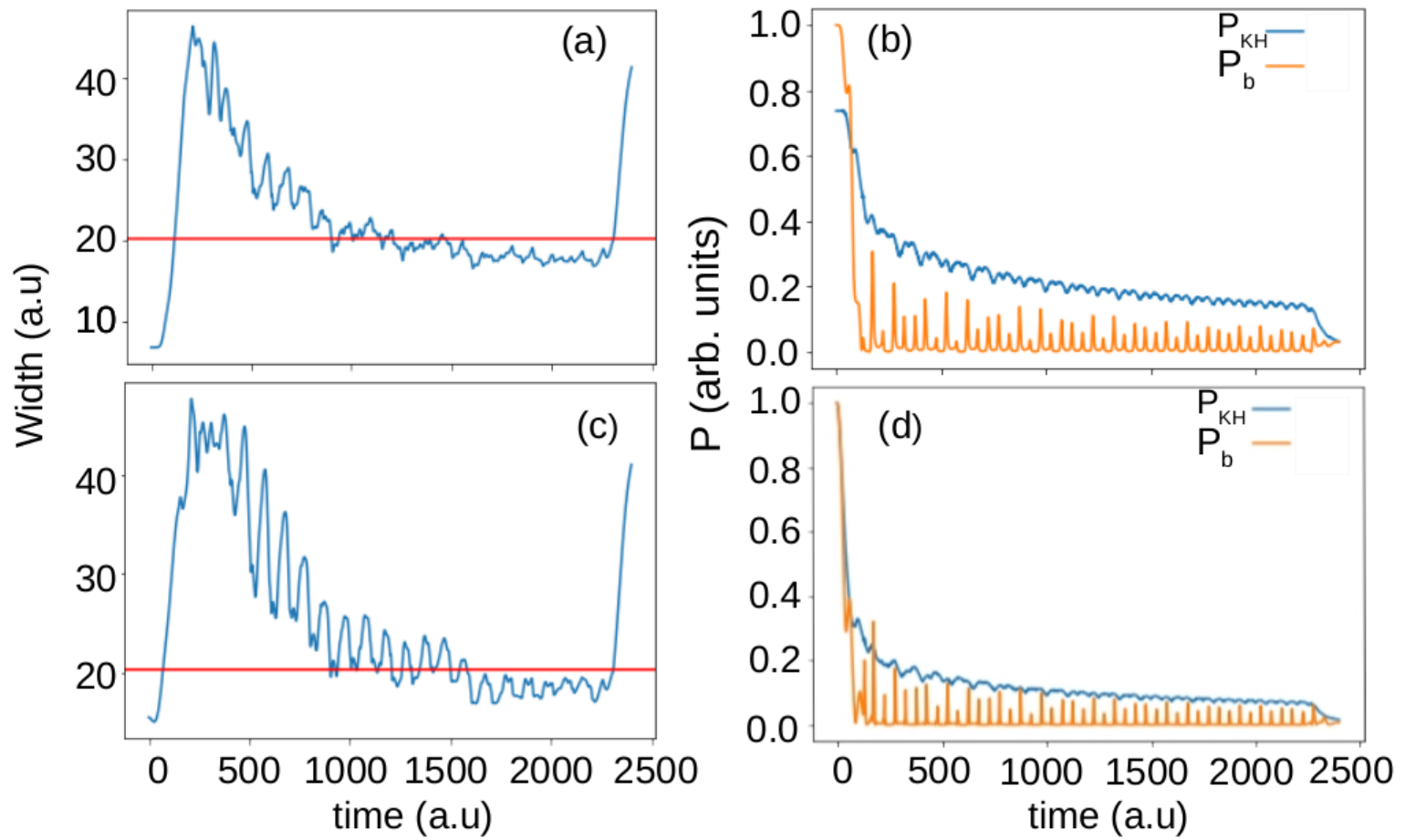}
    \caption{Time-dependent observables calculated solving the TDSE for the binding potential \eqref{eq:potential} and the electric field in Eq.~\eqref{eq:efield}. Panel (a) shows the width of the trapped part of the wave packet using the atomic ground state as initial condition. The horizontal red line is the value $2\alpha_0$. In panel (b) the blue line is the total population of the two KH eigenstates evaluated from Eq.~\eqref{eq:population_KH}, and the orange line is the population of the initial atomic state evaluated from Eq.~\eqref{eq:population_atomic}. Panels (c) and (d) show the same observables as panels (a) and (b), respectively, but considering the first eigenstate of the KH time-averaged potential as the initial condition. All the calculations were performed for a field intensity $I=5.7\times 10^{13} \rm \ W/cm^2$ and frequency $\omega=0.0628 \rm \ a.u$. }
    \label{fig:2}
\end{figure}

\begin{figure}[hbt!]
    \hspace*{-0.8cm}
    \includegraphics[width=1\linewidth]{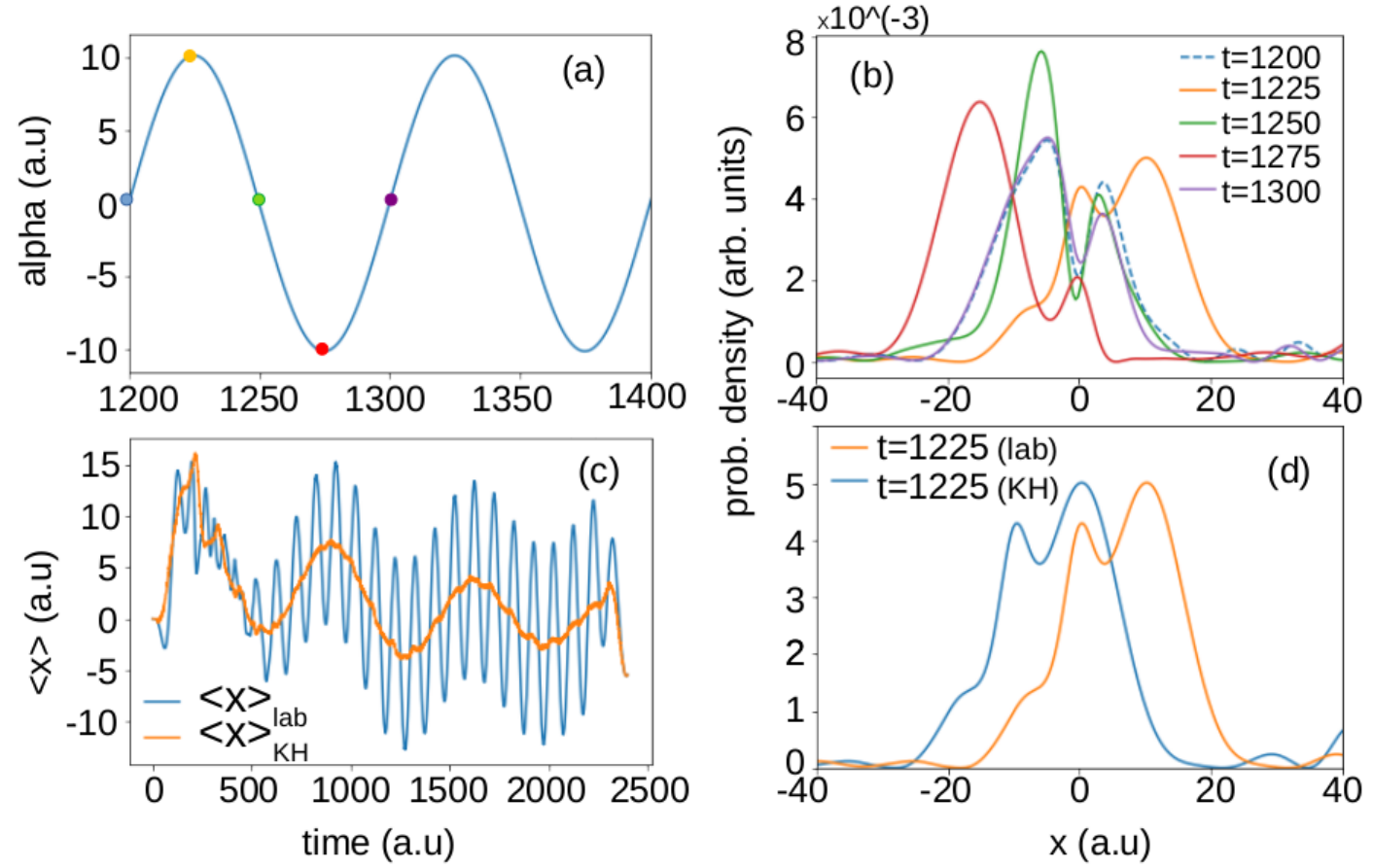}
    \caption{Electron displacement computed classically and from the expectation value of the position operator for the same field parameters as in Fig.~\ref{fig:2}, together with the time-dependent probability density. 
    In panel (a),  we have plotted the electron classical displacement $\alpha(t)$ for two periods of the field in the constant ramp region \eqref{eq:envelope}. Panel (b) shows snapshots of the probability density in the laboratory frame at the times represented by the color points from panel (a) using as the initial condition for the TDSE the first eigenstate of the KH time-average potential.  Panel (c) has the average value of the position of the wave packet in the lab (blue line) and in the KH frame (orange line), again considering the first eigenstate of the KH time average potential as the initial condition. Panel (d) shows the probability density at $t=1225 \rm \ a.u $ as given by the orange line in panel (b) and the same probability density in the laboratory frame, showing the effect of the transformation to the KH frame [Eq.~\eqref{eq:psiKH}].}
    \label{fig:2b}
\end{figure}

In this section, we analyze different observables to determine the validity of the KH regime and whether the KH eigenstates represent a good basis to model the time-dependent dynamics. As a first case study, we consider the propagation of the ground state of the short-range potential [Eq.~\eqref{eq:potential}] under the action of an electric field of the form \eqref{eq:efield} by solving the full TDSE [Eq.~\eqref{eq:tdse}]. Previous studies \cite{popov1999applicability} have shown that in the high-frequency limit, when $\omega\gg|E_g|$, the KH approximation should be valid for any intensity of the laser field. For the parameters used in this work, where $\omega= 0.0628 \rm \ a.u$ and $|E_g|=0.0276 \rm \ a.u$,  thus satisfying the condition $\omega > |E_g|$. Additionally, our parameters satisfiy the Gavrila-Kaminsky condition \cite{Gavrila1984,Richter2016}, $\omega \gg |E^{KH}_0|$, where $E^{KH}_0=-0.01098\rm\ a.u$ is the ground state of the KH time-averaged potential. The present proof of concept builds upon the standard tests performed in order to determine whether the KH regime has been reached. 

Fig.~\ref{fig:2} displays the width of the wave packet and different bound-state populations, which are the most common observables investigated in this context.
The top panels show the results after solving the TDSE [Eq.~\eqref{eq:tdse}] using the ground state of the bound potential as initial condition. In Fig~\ref{fig:2}(a), we plot the width of the trapped part of the wave packet as a function of time. We use a spatial filter and consider that for $-40 \hspace{0.1cm}\mathrm{a.u.}\leq x\leq 40 \hspace*{0.1cm}\mathrm{a.u.}$ the wave packet is trapped, and compute the width defined as twice the standard deviation $\sigma_x$ of the distribution of points within this spatial region. Evidence of reaching the KH regime is given by constant width. The KH regime is only valid during the constant part of the electric field, which starts after a two-cycle turn-on and extends up to $t=2200 \hspace*{0.1cm}\mathrm{a.u}$. The width remains constant, reaching the value of $2\alpha_0$ as discussed in \cite{grobe1993wavepacket}. Further 
evidence of the KH regime is given by the time-dependent population of the KH eigenstates \cite{popov1999applicability}. In Fig~\ref{fig:2}(b) we show how the population $P_b(t)$ of the initial ground state of the short-range potential evaluated from Eq.~\eqref{eq:population_atomic} decays rapidly, while the total population $P_{KH}(t)$ of the KH eigenstates, evaluated from Eq~\eqref{eq:population_KH}, summing over the two eigenstates, remains approximately stationary for the entire time the field amplitude is constant, after a sharp decay occurring during the field turn on. In Figs.~\ref{fig:2}(c) and (d) we plot the same observables as in panels (a) and (b), respectively, but considering the ground state $\phi_0^{KH}(x)$ of the KH time-averaged potential as an initial condition to solve the TDSE [Eq.~\eqref{eq:tdse}]. We can see the same overall behavior. However, the width of the time-dependent wave packet [Fig~\ref{fig:2}(c)] is narrower and the population of the KH eigenstates [blue line in Fig~\ref{fig:2}(d)] is lower compared to its counterpart in Fig~\ref{fig:2}(b).

In Fig.~\ref{fig:2b}, we investigate the electron's classical displacement and the expectation value of the position operator, compared with the time-dependent probability density in the laboratory and KH frames. Because we are interested in the part of the wave packet trapped by the KH potential, the expectation value of the position operator is calculated for the same region in space, that is,  $-40 \hspace{0.1cm}\mathrm{a.u.}\leq x\leq 40 \hspace*{0.1cm}\mathrm{a.u.}$. It is worth noticing that within this region the probability density drops before the onset of stabilization due to ionization. Hence, to define the expectation value we need to enforce normalization of the trapped part of the wave packet to make up for irreversible ionization. The expectation value of the position operator is then defined as
\begin{equation}
  \langle x(t) \rangle = \dfrac{\int_{-40}^{40} |\psi(x,t)|^2x dx}{\int_{-40}^{40} |\psi(x,t)|^2 dx}.
    \label{eq:exp_x}
\end{equation}

Fig.~\ref{fig:2b}(a) shows two periods of the electron classical displacement $\alpha(t)$ for a time where the field strength is constant. This displacement exhibits the expected oscillatory behavior and follows the field closely. 
In  Fig~\ref{fig:2b}(b), we show snapshots of the probability density in the laboratory frame using the first eigenstate as the initial condition in the full TDSE propagation [Eq.~\eqref{eq:tdse}], at the times represented by the color points in Fig~\ref{fig:2b}(a).  We can see how the probability density exhibits the dichotomous structure typical of the KH regime \cite{grobe1993wavepacket} and also follows the field in a way similar to the displacement $\alpha(t)$.  For instance, at times corresponding to a field zero crossing [see the blue, green and purple dots in Fig.~\ref{fig:2b}(a)], the two peaks in the corresponding probability densities [blue, green and purple curves in Fig.~\ref{fig:2b}(b)] are approximately symmetric about the origin, while, for the times matching the field extrema [yellow and red dots in Fig.~\ref{fig:2b}(a)] the probability density is strongly skewed to the left or right [see yellow and red curves in Fig.~\ref{fig:2b}(b)]. In Fig.~\ref{fig:2b}(c) we plot the expectation value of the position operator as defined in Eq.~\eqref{eq:exp_x}. The expectation value of the trapped part of the wave packet oscillates approximately between $[-10:10]\rm \ a.u$ following the field oscillations in the lab frame (blue line).  Additionally, in Figs.~\ref{fig:2b}(c) and (d) we show the effect of transforming from the lab to the KH frame applying Eq.~\eqref{eq:psiKH} to the probability distribution and the average value of the position of the wave packet. The gauge transformation removes the shifts observed in panel (b), so that the probability density in the KH frame no longer follows the field, but still exhibits the same shape and dichotomous structure [Fig.~\ref{fig:2}(d)]. 
Similarly, comparing $\langle x \rangle$ in the KH and lab frames [Fig.~\ref{fig:2}(c); orange and blue curves, respectively], one sees that transforming to the KH frame removes the field oscillations to a great extent, but keeps the behavior that develops in a longer timespan. 

\section{Phase-space analysis and stability}

\begin{figure}[hbt!]
    \centering
    \includegraphics[width=1.09\linewidth]{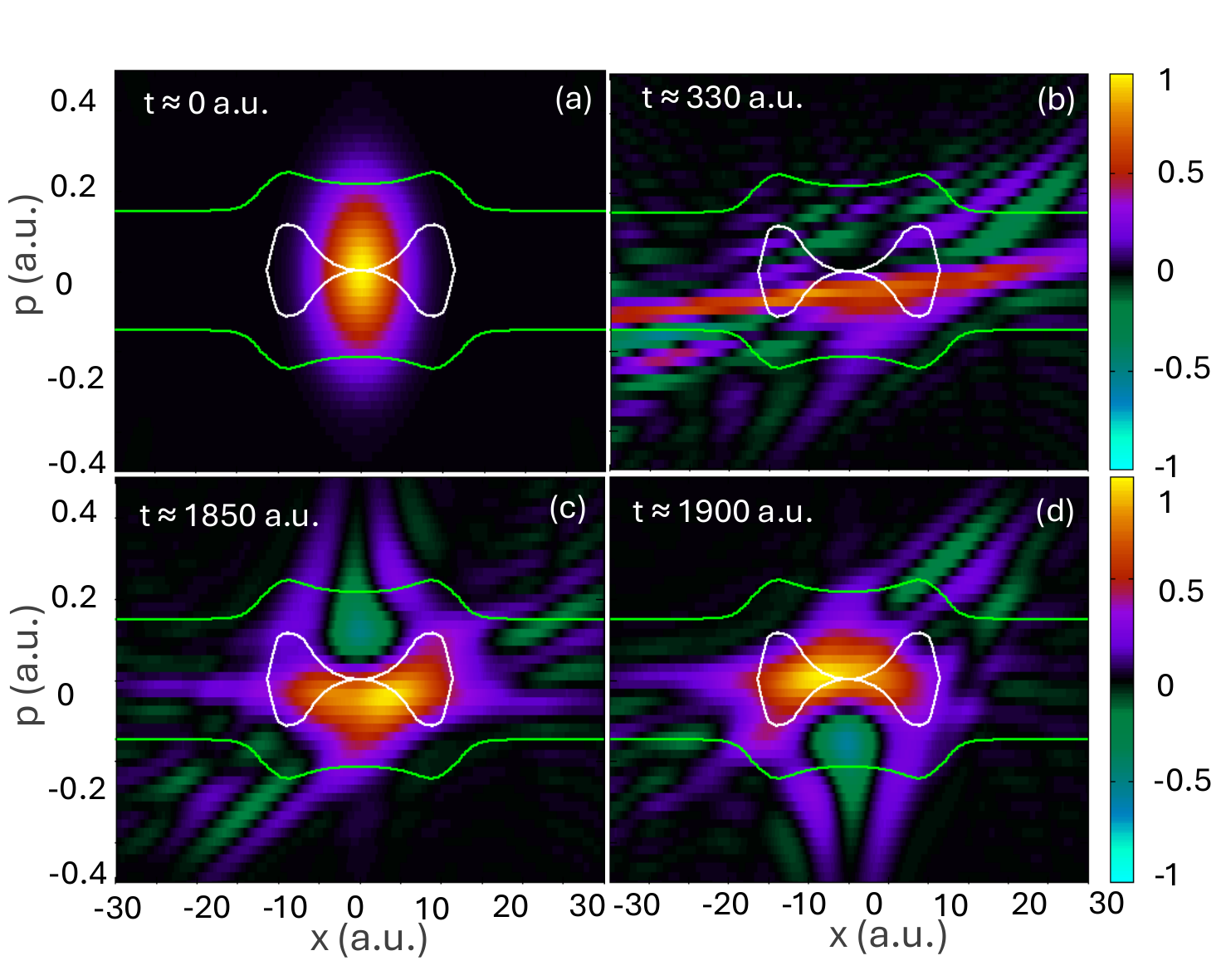}
    \caption{Wigner quasiprobability distributions using the atomic ground state as the initial condition computed in the KH reference frame for the full expression of the potential and the same field parameters as in the previous figure. In Fig.~\ref{fig:3b}(a) the Wigner distribution at $t=0$ is shown. Fig.~\ref{fig:3b} (b) shows the Wigner function at $t=330$ a.u. which is during the constant part of the electric field \eqref{eq:envelope} but when stabilisation has not yet set in and Figs.~\ref{fig:3b}(c) and (d) show the cyclic behavior of the trapped wavepacket at $t=1850$ a.u. and $t=1900$ a.u., respectively, which is after the onset of stabilisation. The thin
white and green lines in the figure give the separatrix at energy $E_{\mathrm{sep}}=-0.0115$ a.u. and the equienergy curve corresponding to $E=0.0125$ a.u., respectively. We have normalized the probability densities to their maximal value in each panel.}
    \label{fig:3b}
\end{figure}

The Wigner quasiprobability distributions [Eq.~\eqref{eq:wignerKH}] mirror the  behavior observed in Figs.~\ref{fig:2} and \ref{fig:2b}, showing evidence of stabilization. In Fig.~\ref{fig:3b}, we display the Wigner quasiprobability distributions calculated assuming that the initial state is the ground state of the atomic potential, as shown in Eq.~\eqref{eq:potential}. These quasiprobability densities are in the KH reference frame. The times have been selected by analyzing Fig.~\ref{fig:2}(b) to find relevant moments of the dynamics, aiming to demonstrate the onset of the KH atom in phase space. The initial Wigner distribution, shown in Fig.~\ref{fig:3b}(a) for $t=0$, is centered at the origin, as expected for the atomic ground state. Subsequently, Fig.~\ref{fig:3b}(b) shows the Wigner function at $t=330$ a.u., which corresponds to a transient region where stabilization has not set in. The original shape of the quasiprobability density is distorted and strong fringes are present around it. Figs.~\ref{fig:3b}(c) and (d) show the quasiprobability flow at $t=1850$ a.u. and $t=1900$ a.u., respectively, which is after the onset of stabilization. This onset is indicated by the emergence of the dichotomous structure in the probability density which led to the observed constant width of the trapped wavepacket shown in Fig.~\ref{fig:2}. The Wigner function elucidates the constant width of the wavepacket in phase space and appears to exhibit a cyclic behavior around the separatrix. The flow associated with this cyclic behavior stays within the green equienergy curve. There are also strong fringes/tails escaping the boundaries of the equienergy curve.  The quasiprobabilities in panels (c) and (d) are very similar except for a $p \rightarrow -p$ reflection. In order to understand this behavior better, we must inspect the KH atom in phase space.

\label{sec:phase-space}

\subsection{Characterization of the KH eigenstates in phase space}
Next, we characterize the KH time-averaged potential in phase space. Although a symmetric coherent superposition of KH eigenstates has been shown in Ref.~\cite{Watson1995}, one must assess the Wigner quasiprobability distribution associated with each eigenstate and how coherent superpositions thereof evolve. We also investigate how the Wigner functions behave regarding classical constraints such as equienergy curves given by Eq.~\eqref{eq:equienergy}. In previous work,  we have found them to provide insight into the dynamics of different quantum pathways \cite{Zagoya2014,Chomet2019,Kufel2020,Chomet2022}. 
\begin{figure}[hbt!]
    \centering
    \includegraphics[width=0.9\linewidth]{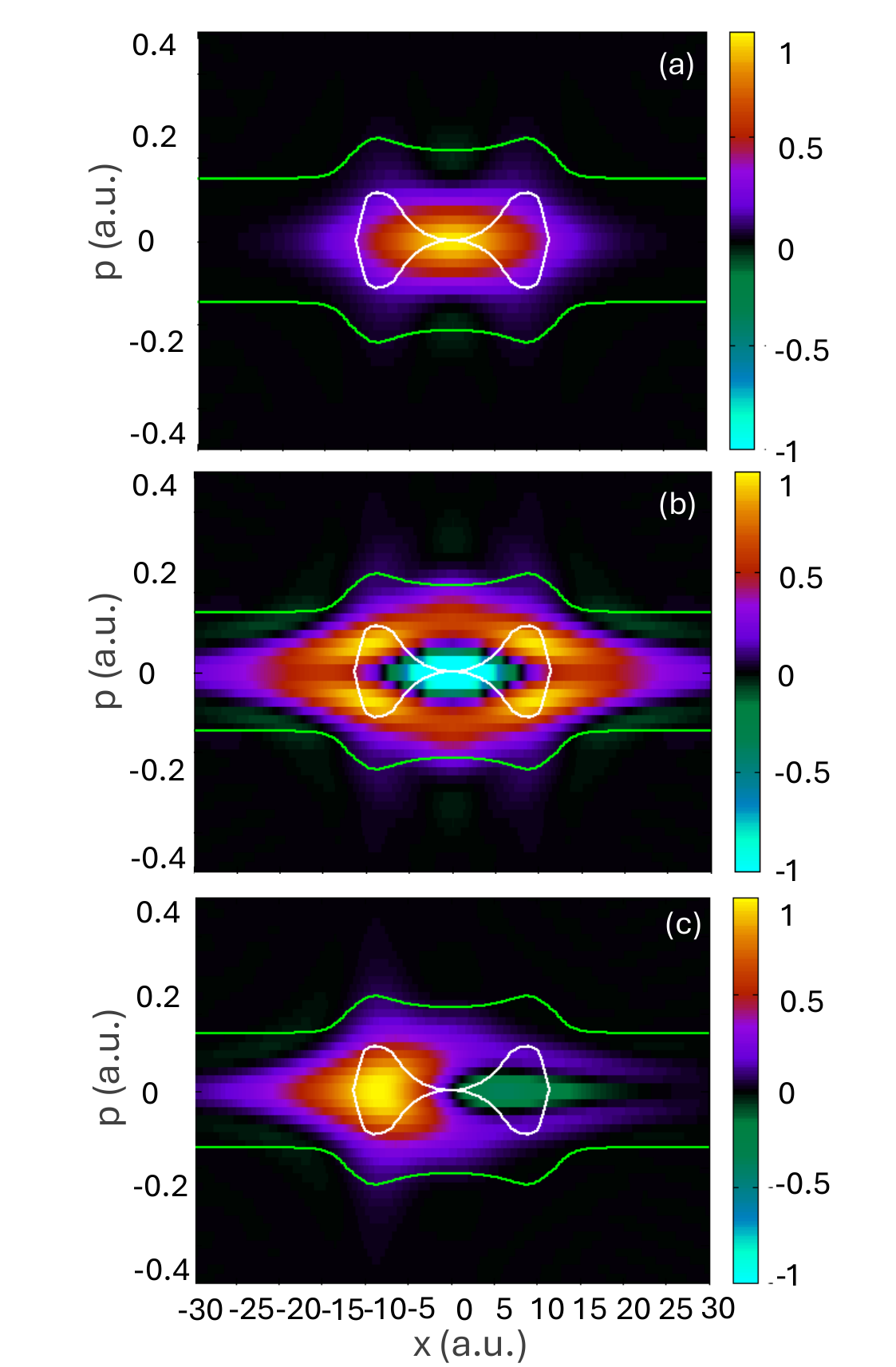}
    \caption{Wigner quasiprobability distributions for the KH eigenstates and their coherent superposition, computed for the time-averaged KH Hamiltonian and the same field parameters as in the previous figure. Panel (a) displays the Wigner quasiprobability distribution for the ground state $\phi_0^{KH}$ of the KH time-averaged potential [Eq.~\eqref{eq:fourierterm}]. Panel (b) presents Wigner quasiprobability distribution for the excited state of the KH time-averaged potential $\phi_1^{KH}$ each of them. Panel (c) depicts the Wigner quasiprobability distribution of the coherent superposition at $t=0$. The thin
white and green lines in the figure give the separatrix at energy $E_{\mathrm{sep}}=-0.0115$ a.u. and the equienergy curve corresponding to $E=0.0125$ a.u., respectively. We have normalized the probability densities to their maximal value in each panel.}
    \label{fig:3}
\end{figure}

In Figs.~\ref{fig:3}(a) and (b), we show the Wigner distribution \eqref{eq:wignerKH} for the ground and excited eigenstates of the KH time-averaged potential, respectively, together with the separatrix at $E_{\mathrm{sep}}=-0.0115$ a.u. and the equienergy curve for $E=0.0125$ a.u. (white and green curves, respectively). The corresponding probability densities are plotted in Fig.~\ref{fig:2b}. 
Overall, the Wigner functions associated with the stationary states are roughly contained by the outermost equienergy curve in the figure, although there is a position and momentum spread as required by the uncertainty relation. For the ground KH eigenstate [Fig.~\ref{fig:3}(a)], the Wigner quasiprobability distribution is peaked at the origin, and its width extends beyond the wells determined by the separatrix. In contrast, for the excited KH eigenstate [Fig.~\ref{fig:3}(b)], the Wigner function appears to peak at the turning points defined by the separatrix and to extend beyond, roughly following the shape of the external, green equienergy curve. The Wigner function is also negative at the origin, which is evidence of nonclassical behavior. Since these are the eigenstates of the KH time-averaged potential, 
we expect the probability density to remain stationary. 

 Additionally, we build an equally weighted coherent superposition of the two eigenstates, given by
\begin{equation}\label{eq:coherent}
\ket{\psi_{\mathrm{coh}}}=\dfrac{1}{\sqrt{2}}(\ket{\phi^{KH}_{0}}+\ket{\phi^{KH}_{1}}),
\end{equation}
and localize the initial wave function in the left well of the KH time-averaged potential. The Wigner quasiprobability distribution corresponding to the coherent superposition in Eq. \eqref{eq:coherent}  is shown in Fig.~\ref{fig:3}(c) for $t=0$.  This coherent superposition gives a quasiprobability density with a positive peak on the left well, and a weaker, negative peak on the right well. The momentum ranges remain confined by the external equienergy curve. A similar shape is observed for a diatomic molecule, if a localized coherent superposition of the ground and excited eigenstates is constructed \cite{Chomet2019}. 

\begin{figure}[hbt!]
    \centering
    \includegraphics[width=1\linewidth]{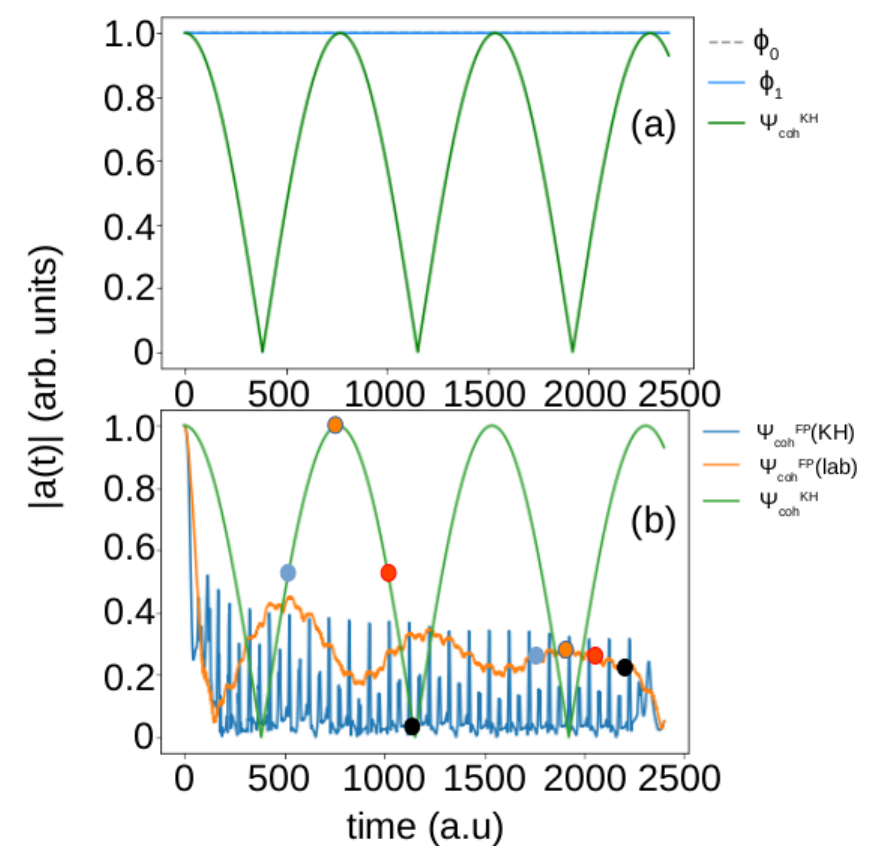}
    \caption{Absolute value of the autocorrelation function $|a(t)|$ computed for different initial states considering the full and time-averaged KH Hamiltonians and the same field and grid parameters as in the previous figures. The superscript in the legends indicate the full potential (FP) and the time-averaged KH (KH) Hamiltonians. Panel (a) shows $|a(t)|$ considering three different initial conditions in the KH time-averaged potential: the first $\phi_0$ and second eigenstates $\phi_1$ individually (shown with a dashed gray and blue lines respectively), and the coherent superposition of the two eigenstates $\Psi_{\rm coh}$ (oscillatory green line). Panel (b) shows $|a(t)|$ using the coherent superposition as the initial condition in the full Hamiltonian, calculated both in the laboratory frame (blue line) and in the KH frame (orange line), for comparison we add the results from panel (a) (green line). The points indicate the times for the maximum, minimum and midpoint values of $|a(t)|$. The blue and red dots signpost the times for which the probability density is spread across both wells and correspond to $t\approx 578$ $ \rm a.u.$ and $t\approx 963 $ $ \rm a.u.$, respectively on the green line and $t\approx 1695$  $ \rm a.u.$ and $t\approx 2050$  $ \rm a.u.$, respectively on the orange line. The orange dots indicate the times for which the probability density is localized on the left well, which are around $t\approx 770$ $ \rm a.u.$ and $t\approx 1900 $ $ \rm a.u.$  for the time-averaged (green line) and the full potential (orange line), respectively. The black dots highlight the times for which the probability density is localized on the right well, which are around $t\approx 1155$ $ \rm a.u.$ and $t\approx 2196$ $ \rm a.u.$ for the time-averaged (green line) and the full potential (orange line), respectively. One should note that the dots in the green and orange curves indicate similar behaviours, but are chosen for different times. This takes into account due to transient effects for the case of the full expression.}
    \label{fig:4}
\end{figure}

If the coherent superposition \eqref{eq:coherent} is chosen as the initial condition in Eq.~\eqref{eq:Kh average} we expect the wave packet to oscillate with a frequency $\omega_{10}=|E^{KH}_1-E^{KH}_0|=8.16 \times 10^{-3}$ a.u., which gives a period of $T_{10}=770$ a.u., which is almost eight times a field cycle. This behavior can be inferred from the evolution of the wave packet  $\psi_{coh}(x,0)=\bra{x}\ket{\psi_{coh}}$, which is computed using the time evolution operator
\begin{equation}
U^{(0)}_{KH}(t,t_0)=\exp{-iH_{KH}^{(0)}(t-t_0)},
\label{eq:tevolKH}
\end{equation}
where $H_{KH}^{(0)}$ s the time-averaged KH Hamiltonian defined by Eq.~\eqref{eq:Kh average} and we choose $t_0=0$ for simplicity. This leads to the probability density
\begin{eqnarray}
|\psi_{\mathrm{coh}}(x,t)|^2&=&\frac{1}{2}\left[|\phi^{KH}_{0}(x)|^2 + |\phi^{KH}_{1}(x)|^2\right]\\&+&\mathrm{Re}[\phi^{KH}_{1}(x)(\phi^{KH}_{0}(x))^*]\cos{\omega_{10}t}. \notag
\end{eqnarray} 
The same time dependence is encountered for the Wigner quasiprobability distribution in the KH frame, which is calculated from Eq.~\eqref{eq:wignerKH} using the time evolution operator \eqref{eq:tevolKH}. Explicitly, this gives
\begin{eqnarray}
    W_{\mathrm{coh}}^{KH}(x,p,t)|^2&=&\frac{1}{2}\left[| W_{1}^{KH}(x,p)+W_{0}^{KH}(x,p)\right] \notag\\&+&\mathrm{Re}[W_{10}^{KH}(x,p)]\cos{\omega_{10}t},
\end{eqnarray}
with 
\begin{equation}
    W_{n}^{KH}(x,p)=\frac{1}{2\pi}\int^{\infty}_{-\infty}\hspace*{-0.3cm}e^{ip\xi}\phi^{KH}_n(x-\xi/2)(\phi^{KH}_n(x+\xi/2))^*d\xi,
\end{equation}
$n=0,1$, and
\begin{equation}
    W_{10}^{KH}(x,p)=\frac{1}{2\pi}\int^{\infty}_{-\infty}\hspace*{-0.3cm}e^{ip\xi}\phi^{KH}_1(x-\xi/2)(\phi^{KH}_0(x+\xi/2))^*d\xi.
\end{equation}
In Fig.~\ref{fig:4}a) we show the absolute value of the autocorrelation function $a(t)$ defined as

\begin{equation}\label{eq:auto}
  |a(t)|=\left|\int \psi(x,t)^*\psi(x,0) dx\right|,
\end{equation}

\noindent evaluated after solving the TDSE Eq.~\eqref{eq:Kh average} for the time-averaged KH potential using the two eigenstates separately as initial conditions (superimposed dashed gray and blue lines in the figure) and their equally weighted coherent superposition as defined in Eq.~\eqref{eq:coherent} (green line in the figure). The autocorrelation functions change periodically with frequency $\omega_{10}$  indicating how the wave packet oscillates periodically between the two wells of the KH time-averaged potential.  

Subsequently, we study the time-dependent dynamics of the initial state given by the coherent superposition in Eq.~\eqref{eq:coherent} considering the full potential and electric field. We solve the TDSE Eq.~\eqref{eq:tdse} in the lab frame and by applying the transformation \eqref{eq:psiKH} we analyze the observables and obtain the Wigner quasiprobability distributions in the KH frame. Fig.~\ref{fig:4}(b) shows the absolute values of the autocorrelation function for this study in the lab frame (blue line) and the KH reference frame (orange line). As a reference we have included the calculations done for the KH time-averaged potential shown with the green lines in both panels in Fig.~\ref{fig:4}. By comparing the results in the KH frame (orange line) with those computed using the KH time-averaged potential (green line), we can see how by transforming to the KH frame the main oscillations obtained in the KH time-averaged potential, which are associated with the cyclic behavior of the Wigner functions, are present.  However, the frequency of the oscillations is different. This could be due to the contribution of the time-dependent terms of the expansion or transient effects introduced during the turn-on ramp of the laser field. Nonetheless, the main features of the KH regime are present, with the wave packet moving periodically from one well to the other. 
A periodic behavior is also observed for a field-free diatomic molecule, for much smaller internuclear separation than the distance between the wells of the KH potential \cite{Chomet2019,Kufel2020}. However, the physical mechanisms underpinning this behavior differ, as we discuss next.

\subsection{Phase-space behavior of coherent superpositions of eigenstates}

\begin{figure*}[hbt!]
    \centering
    \includegraphics[width=1\linewidth]{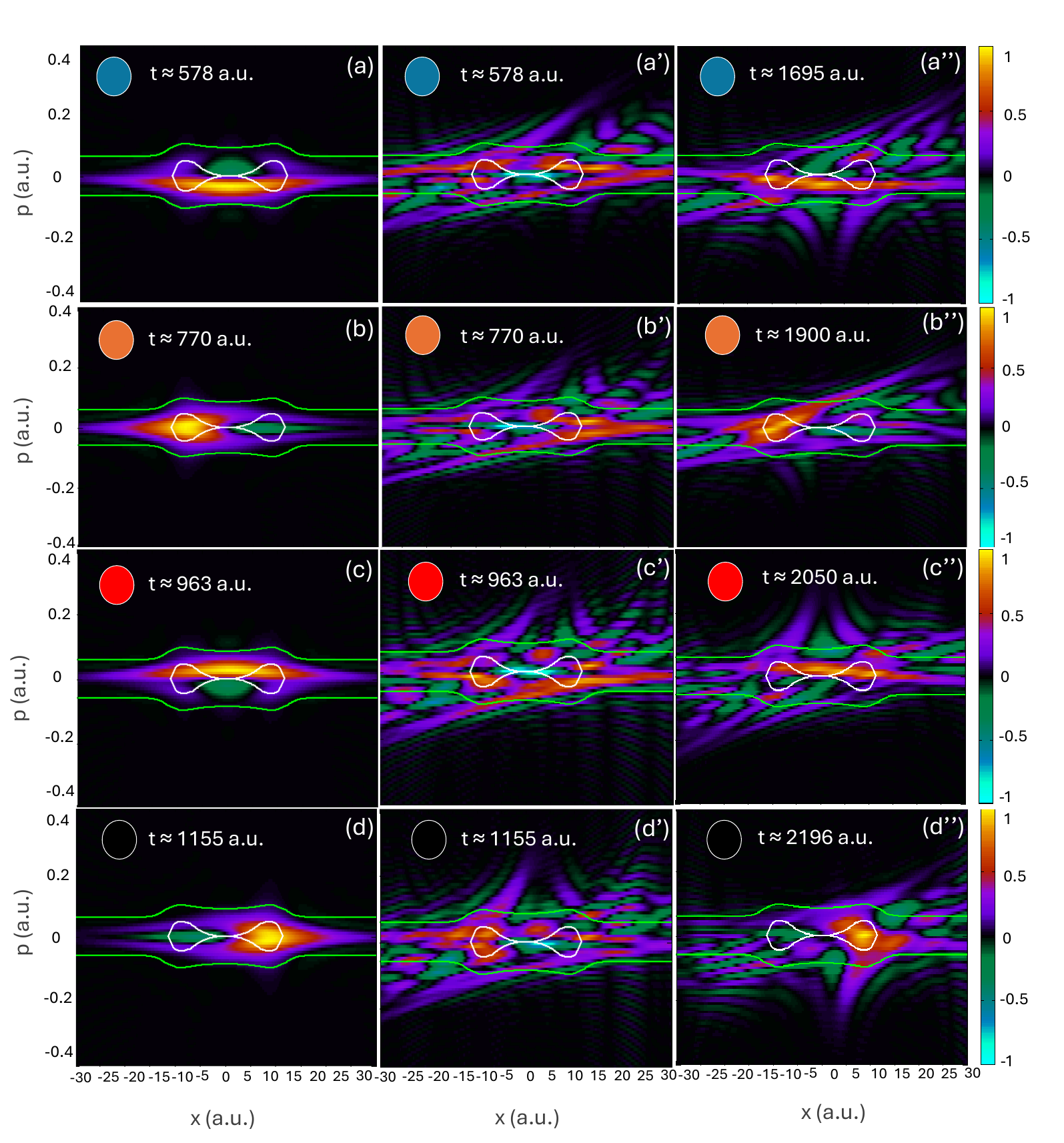}
    \caption{ Wigner quasiprobability distribution at different times corresponding to the points indicated in Fig.~\ref{fig:4} using the coherent superposition of the two eigenstates as the initial condition for the KH time-averaged potential (left column) and the full expression of the potential (middle and right columns) and the same field parameters as in the previous figures. The circles on the left upper corner of the figure employ the same color as in Fig.~\ref{fig:4}, in order to indicate the times marked the dots in the autocorrelation functions (ACFs). Those in the left column are associated with the ACF computed for the time-averaged potential, while those on the middle and right column are indicated in the full ACF (green and orange curves in Fig.~\ref{fig:4}, respectively). The temporal snapshots are given from top to bottom. Panels (a), (a') have been calculated for $t \approx 578 \rm a.u.$ and (a'') for $t \approx 1695 \rm a.u.$ respectively, panels (b), (b') have been calculated for $t \approx 770 \rm a.u.$ and (b'') for $t \approx 1900 \rm a.u.$, respectively, panels (c), (c') have been calculated for $t \approx 963 \rm a.u.$ and (c'') for $t \approx 2050 \rm a.u.$, respectively, and panels (d), d') have been calculated for $t \approx 1155 \rm a.u.$ and d'') for $t \approx 2196 \rm a.u.$, respectively. The thin
white and green lines in the figure give the separatrix at $E_{sep}=-0.0115$ a.u. and the equienergy curve corresponding to $E=0.0125$ a.u., respectively. We have normalized the probability densities to their maximal value in each panel.}
    \label{fig:5}
\end{figure*}

Here, we compare the phase-space behavior in the full and KH time-averaged potential at relevant times of the dynamics given by the color points in Fig.~\ref{fig:4}(b), which represent maxima, minima and mid-points of the autocorrelation functions computed using the time-averaged and full KH Hamiltonian. The times associated with the mid-point of the autocorrelation function for the time-averaged potential correspond to a probability density equally distributed across the two wells of the KH averaged potential. These instances are signposted with blue and red dots for the time-averaged and the full potential, although these behaviors occur at different times. The orange (black) dots indicate the times for which the wave packet is localized on the left (right) wells. The corresponding Wigner quasiprobability distributions are plotted in Fig.~\ref{fig:5}, for the time-averaged KH dynamics [left column] and full temporal evolution [middle and right columns]. 

In the left column of Fig.~\ref{fig:5} we show snapshots of the Wigner quasiprobability function evaluated after solving the TDSE [Eq.~\eqref{eq:Kh average}] using the equally weighted coherent superposition $\psi_{\mathrm{coh}}(x,0)$ of the two KH eigenstates as the initial condition for the KH time-averaged potential. The specific times chosen for those snapshots match the dots in the green curve in Fig.~\ref{fig:4}(b). For the time-averaged potential, the Wigner quasiprobability distributions exhibit a cyclic behavior matching the oscillations of the autocorrelation function, without any significant spill outside the region of interest. The quasiprobability flow extends spatially to over 30 a.u. and oscillates between the wells of the dichotomous potential,  spilling beyond the classical separatrix (see white curves in the figure). However, it is bounded to a region of small momenta, within the green equienergy curves. Figs.~\ref{fig:5}(a) and (c) show quasiprobability located mainly in the central region between the two wells, with the Wigner plot in (a) being the mirror image of that in (c) regarding the x-axis. In contrast, for Figs.~\ref{fig:5}(b) and (d) the quasiprobability density is mostly located around the wells, and they are the mirror image of each other about the p-axis. 

The central column of Fig.~\ref{fig:5} exhibits the  Wigner functions at the same instants of time as in the left column, that is, indicated by the colored points in the green curve in Fig.~\ref{fig:4}(b), but computed using the full time-dependent potential. Overall, we observe a lack of agreement in the results from the full and time-averaged potential. Although there is a cyclic motion for the quasiprobability flow corresponding to the full dynamics, the times for which the Wigner functions are located on the left, right and central regions are mismatched. For instance, the positive quasiprobability density in Fig.~\ref{fig:5}(b) [Fig.~\ref{fig:5}(d)] is located around the left (right) well, while its counterpart in Fig.~\ref{fig:5}(b')[Fig.~\ref{fig:5}(d')] is located on the right (left). Similarly, the positive quasiprobablity density in Fig.~\ref{fig:5}(a) [Fig.~\ref{fig:5}(c)] is located in the phase-space region below (above) the saddle determined by the separatrix, while its counterpart in Fig.~\ref{fig:5}(a')[Fig.~\ref{fig:5}(c')] is located above (below) it. Furthermore, the behavior is not purely cyclic, with quasiprobability tails spilling towards higher momentum regions.  

These differences are likely due to transient effects being introduced by the turn-on ramp of the laser field for the full expression of the potential, which is not captured in the time-averaged case. During the turn-on, the field introduces a time-dependent perturbation that distorts and evolves the wave packet over time. By the time the laser field reaches its steady state intensity, the wave function in the full KH potential will have significantly evolved from its initial-state configuration. These transients can cause such discrepancies early on, which in turn delays the time frame at which stabilization is achieved for the full KH potential compared to the time-averaged case. The left column is already operating in the stabilization regime as shown by the trapped nature of the electron wave packet whereas the central column appears to still be in the ionization regime, as indicated by the prominent fringes/leakage that can be attributed to ionization effects. Similar patterns have been identified elsewhere, for the KH atom \cite{Watson1995,Watson1996} and also in the low-frequency regime \cite{Czirjak2000,Zagoya2014,Chomet2019}. Key features of this leakage are that it extends towards higher momenta and the presence of fringes with alternating signs for the quasiprobability flow. These fringes indicate the quantum interference of different ionization events \cite{Czirjak2000}. 

Therefore, it is physically more meaningful to compare the Wigner quasiprobability flow after the onset of stabilization. These results are displayed in the right column of Fig.~\ref{fig:5}, in which we have chosen snapshots matching the colored points in the orange curve in Fig.~\ref{fig:4}(b). For these snapshots, the main dynamics of the system in the time-averaged case are reflected in the full expression.  
Interestingly, the comparison between the left and right columns demonstrates that despite the effect of the turn-on ramp in the beginning, the main features of the KH atom are retained in the Wigner function albeit at a later time interval. In both cases, we can see that the wave packet is essentially trapped in this spatial region, moving from one well to the other. The Wigner function exhibits a clockwise movement, with its period, which can be inferred from the autocorrelation function shown in Fig.~\ref{fig:4}(b), depending on
the well separation. This is somewhat similar to the case of a diatomic molecule, with $2\alpha_0$ in our case playing the role of the internuclear distance. This cyclic motion is roughly in phase for both the time-averaged and full potentials. For instance, in Figs.~\ref{fig:5}(a) and (a'') [Figs.~\ref{fig:5}(c) and (c'')], there is a strong positive quasiprobability density in the central region below (above) the saddle, while, in Figs.~\ref{fig:5}(b) and (b'') [Figs.~\ref{fig:5}(d) and (d'')], they are peaked around the left [right] well. Still, the full results also exhibit ionization, observed as fringes extending the quasiprobability density towards higher momenta and distorting the cyclic motion. Nonetheless, they seem to be less pronounced than those observed before the onset of stabilization. This is very evident comparing Figs.~\ref{fig:5}(d') and (d''), but can also be observed in subtler ways inspecting Figs.~\ref{fig:5}(c') and (c''), or Figs.~\ref{fig:5}(b') and (b'').

\begin{figure}[hbt!]
    \centering
    \includegraphics[width=1\linewidth]{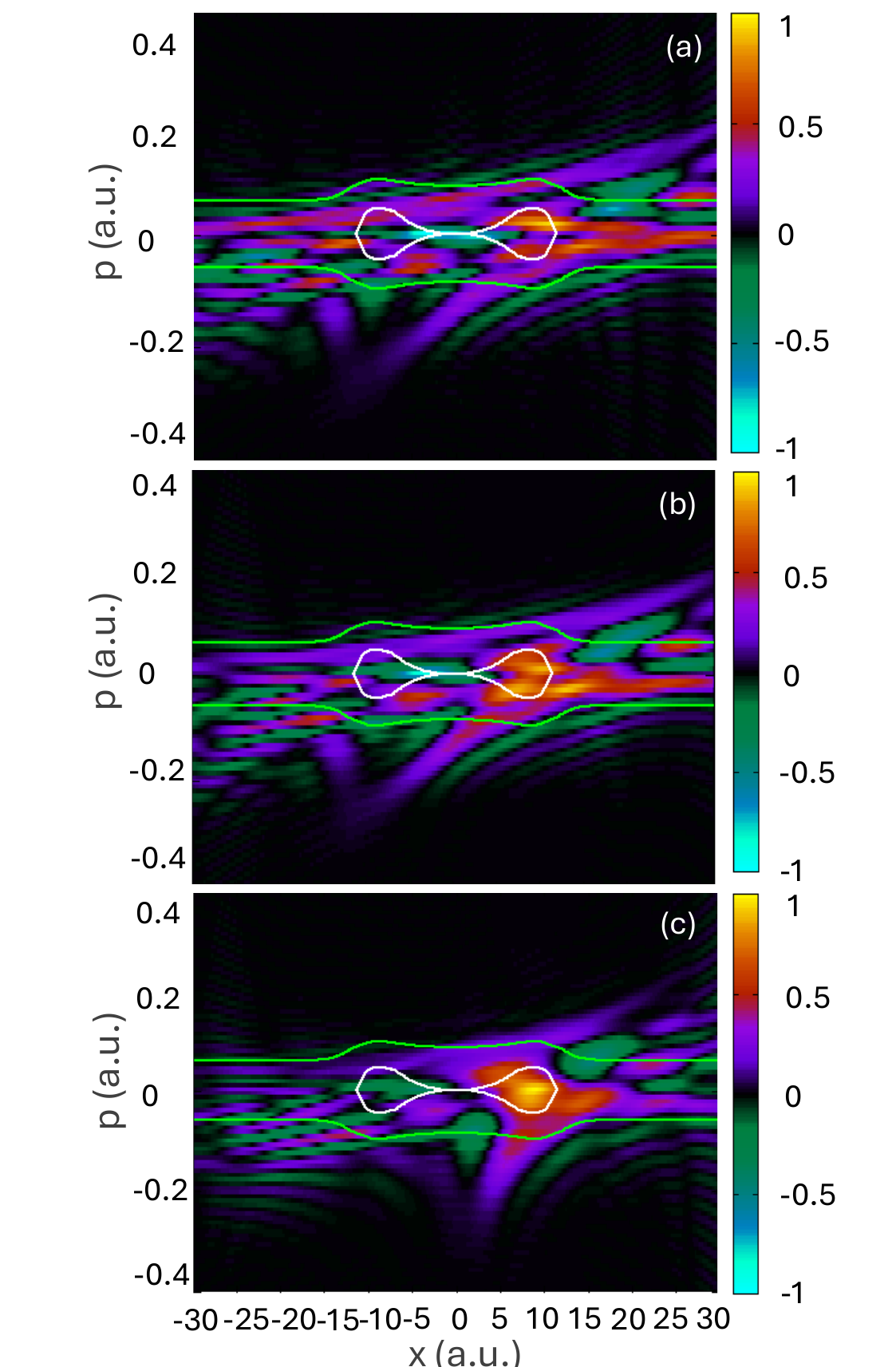}
    \caption{Wigner function at different times using the coherent superposition of the two eigenstates as the initial condition for the full expression of the potential and the same field parameters as in the previous figures. Panels (a), (b) and (c) have been calculated for $t \approx 720$ a.u., $t \approx 1520$ a.u. and $t \approx 2198$ a.u., respectively. The thin white and green lines in the figure give the separatrix and equienergy curve corresponding to $E=0.0125$ a.u., respectively. We have normalized the probability densities to their maximal value in each panel. }
    \label{fig:6}
\end{figure}

In Fig.~\ref{fig:6}, we illustrate the onset of stabilization and its implication on these fringes more systematically, reinforcing the statement that the tail-shaped features are associated with ionization. The snapshots in the figure were chosen such that the Wigner functions exhibit strong peaks around the right well of the KH potential, but considering the fully time-dependent dynamics. The times in Figs.~\ref{fig:6}(a) and (b) were chosen before one and two periods $T_{10}$ associated with the KH eigenstates, and Fig.~\ref{fig:6}(c) was calculated for a time just before the field turn-off. For the three times, the field amplitude is kept constant, meaning they are after the field turn-on and before its turn-off.  The fringes that form a tail towards higher momentum regions become fainter with time as shown by comparing the three panels in Fig.~\ref{fig:6}. This suppression of ionization, along with wave-packet localization in one of the wells, indicates that we have transitioned to the stabilization regime.  

Next, we investigate to what extent the KH approximation \eqref{eq:fourierterm}, characterized by the zeroth order term of the Fourier expansion \eqref{eq:fourier} of the full potential, holds, and how that manifests in the context of our Wigner functions. This will be done by initially using the ground state of the KH atom or the coherent superposition of the two KH eigenstates and allowing them to evolve under the full KH potential up to the point when the system enters the stabilization regime. This instance can be identified by observing the time at which the dichotomous structure emerges in the atomic probability density [see Fig.~\ref{fig:2b} and the ensuing discussion]. Then, we take the evolved wave packet at this time as the initial condition for the time-averaged KH potential. Thus, the starting points of both the time-averaged and full KH potentials in the Wigner function calculation align more closely at the onset of stabilization. 

\begin{figure}[hbt!]
\hspace*{-0.5cm}
    \includegraphics[width=1.15\linewidth]{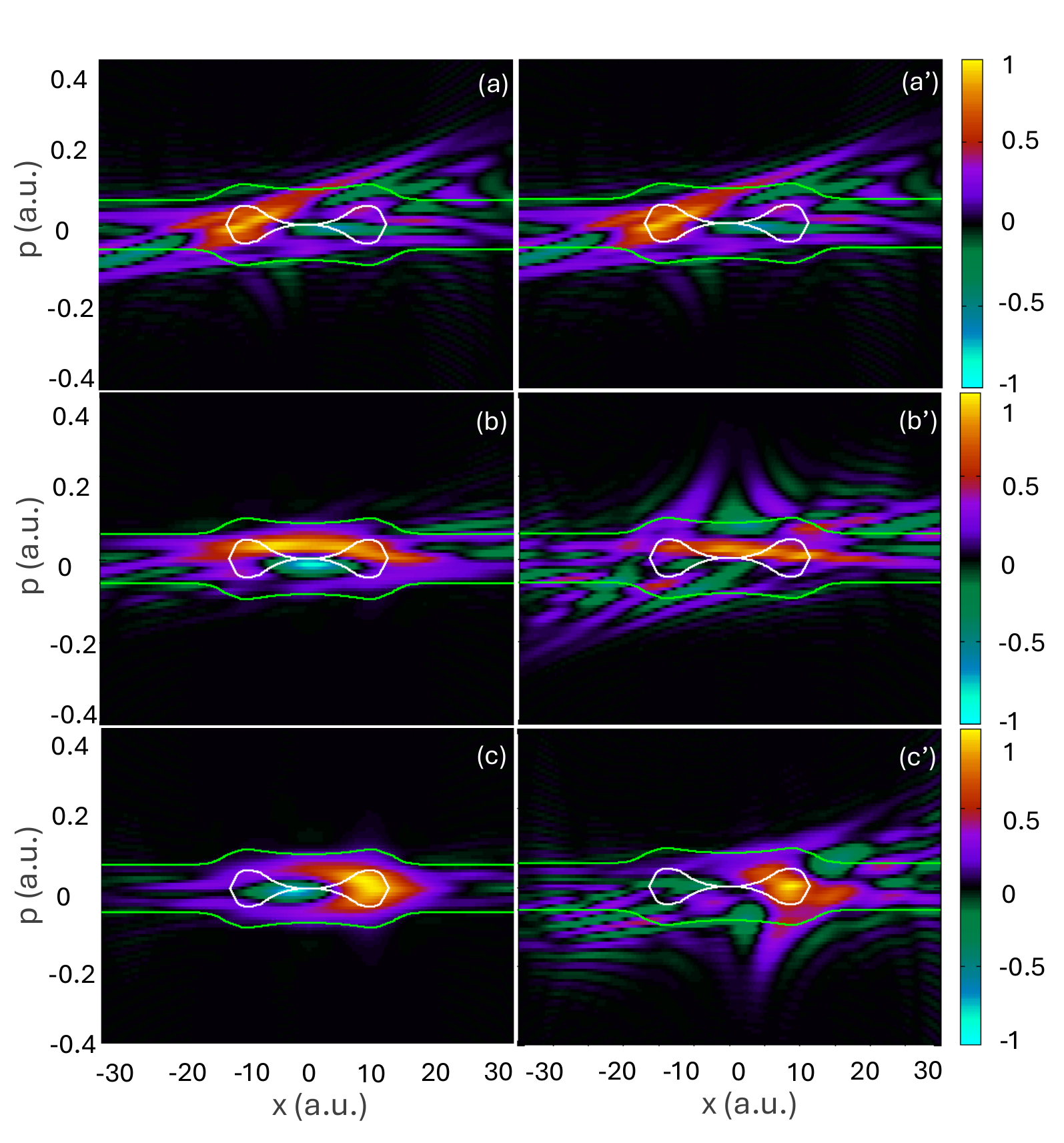}
    \caption{Wigner function at different times using a coherent superposition of KH eigenstates as the initial condition in the full KH potential expression and the same field parameters as in the previous figures. For the case of the KH-time averaged potential (left column) we select the wave packet at $t \approx 1900$ a.u. as its initial condition and for the full KH expression (right column) the initial condition remains the same. The temporal snapshots are given from top to bottom. Panels (a) and (a') have been calculated for  $t \approx 1900$ a.u., panels (b) and (b') have been calculated for  $t \approx 2050$ a.u., respectively, panels (c) and (c') have been calculated for $t \approx 2196$ a.u. The thin
white and green lines in the figure give the separatrix and equienergy curve corresponding to $E=0.0125$ a.u., respectively. We have normalized the probability densities to their maximal value in each panel. }
    \label{fig:7}
\end{figure}

 In Fig.~\ref{fig:7} we show the results of this test for when we initially start off with a coherent superposition of KH eigenstates in the full KH potential expression and select the wave packet at $t \approx 1900$ a.u. as the initial condition for the time-averaged potential. 
 The left and right columns of Fig.~\ref{fig:7} have been computed for the time-averaged and full potential, respectively. The localization of the electron and the cyclic motion of the distribution over time are well described in the time-averaged case when comparing the same times.  However, the tails associated with ionization disappear at later times if the time propagation is performed using the time-averaged potential. This can be seen by comparing Figs.~\ref{fig:7}(b) and (b'), and Figs.~\ref{fig:7}(c) and (c'), and means that ionization takes place even if the stabilization regime has been reached.  

\begin{figure}[hbt!]
    \hspace*{-0.2cm}
    \includegraphics[width=1.09\linewidth]{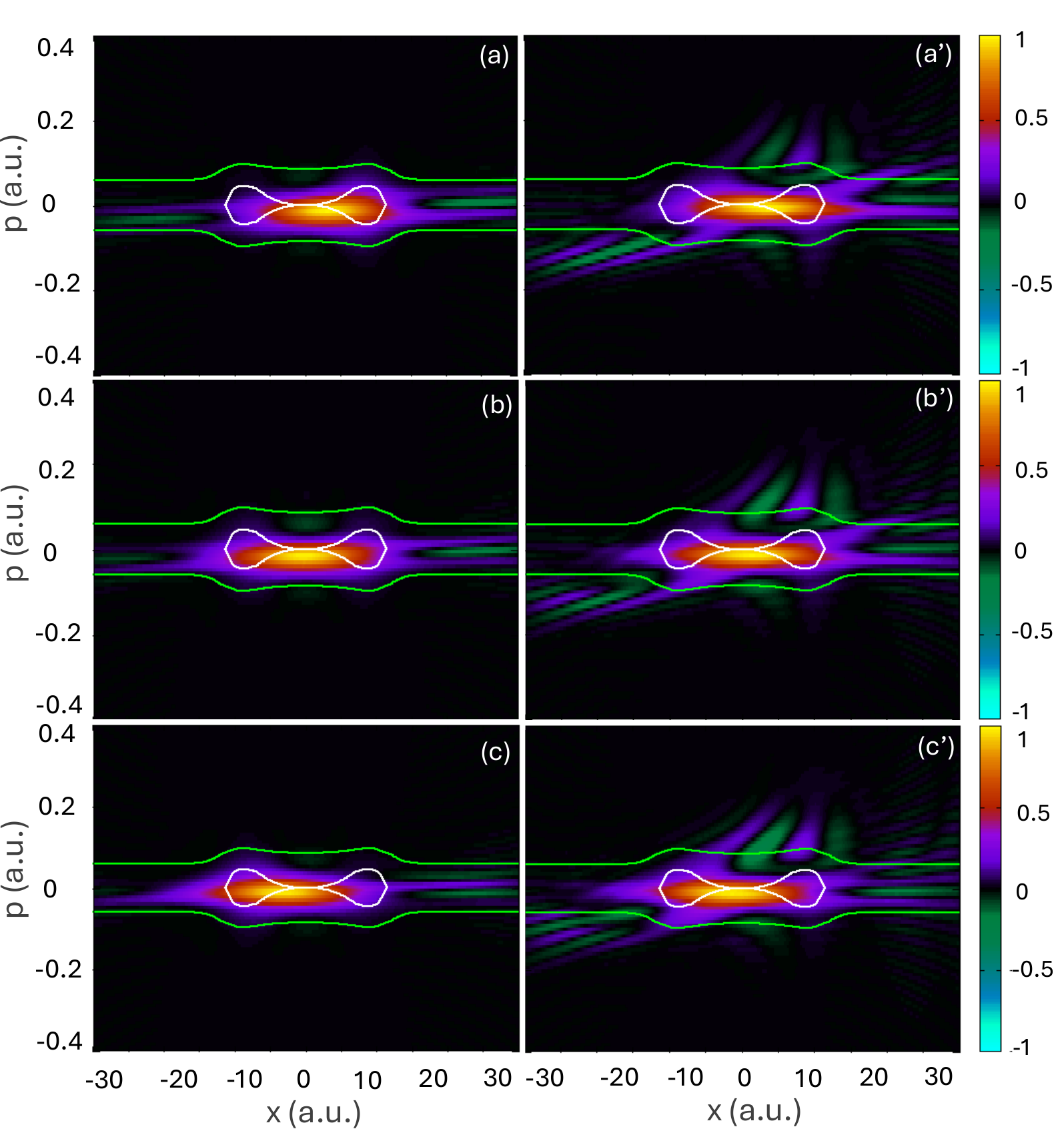}
    \caption{Wigner function at different times computed using the ground state of the KH atom as the initial condition in the full KH potential expression and the same field parameters as in the previous figures. For the case of the KH-time averaged potential (left column) we select the wave packet at $t \approx 1200$ a.u. as its initial condition and for the full KH expression (right column) the initial condition remains the same.The temporal snapshots are given from top to bottom. Panels (a) and (a') have been calculated for  $t \approx 1671$ a.u., panels (b) and (b') have been calculated for  $t \approx 1771$  a.u., respectively, panels (c) and (c') have been calculated for $t \approx 1871$ a.u.. The thin white and green lines in the figure give the separatrix and equienergy curve corresponding to $E=0.0125$ a.u., respectively. We have normalized the probability densities to their maximal value in each panel.}
    \label{fig:8}
\end{figure}

In Fig.~\ref{fig:8}, we illustrate how ionization can be minimized if, instead, we take the ground state of the time-averaged KH potential as the initial condition for the full expression and select the wave packet at $t \approx 1200$ a.u. as the initial condition for the time-averaged potential [left column in Fig.~\ref{fig:8}], as well as for the full dynamics  [right column in Fig.~\ref{fig:8}]. In this case, the Wigner function is approximately stationary, localized near the origin of the phase space, and the dynamics are very similar for the full and the time-averaged potentials. A direct comparison with the previous figure shows that the tails associated with ionization are strongly suppressed. Furthermore, stabilization sets in at an earlier time, making the Wigner comparison even clearer.

\section{Conclusions}
\label{sec:conclusions}
 
In this work, we have investigated the phase-space behavior of the Kramers-Henneberger (KH) atom with emphasis on understanding its key features and its sensitivity to different initial conditions. We find that ionization suppression and the stability of the system are strongly influenced by how these conditions are chosen. For simplicity, we employ the one-dimensional model proposed in \cite{popov1999applicability} as it supports only two KH eigenstates. This facilitates the analysis of what fraction of the bound-state population leaks into the continuum, the main features associated with KH eigenstates, and the validity of the KH approximation. In contrast, using a long-range softcore potential may lead to intricate coherent superpositions of KH states, which are created depending on the pulse shape and may be difficult to control \cite{Watson1996}, or even to phase-space regions for which there is a quasi-continuum, which is then studied using periodic-orbit theory \cite{Sundaram1992,Norman2015,Floriani2024}. 

A detailed study of the time-dependent wave function obtained with the full Hamiltonian shows that there is a time range after the pulse turn-on and before its turn-off, for which its behavior is consistent with the presence of an effective dichotomous potential, such as a double-peaked structure. This agrees with the findings in \cite{popov1999applicability} and is supported by analyzing several observables, such as the width and average position of the wave packet, and the total population of the KH eigenstates. A constant width, a stable population of the KH eigenstates, and an oscillatory average position indicate that stabilization is present. This investigation assures that we are working in the right physical regime, and sets the stage for the subsequent phase-space studies. 

We started by understanding the Wigner quasiprobability distributions associated with the KH eigenstates, and comparing them with the classically determined equienergy curves. For a time-averaged KH potential, these Wigner functions are stationary and have well-defined parities. The ground-state Wigner function is localized mostly at the origin, between the two wells, and is non-negative, while the excited-state Wigner function is peaked around the wells. There is also a large negativity near the origin, which hints at the nonclassical behavior of the Wigner functions in that region. A  common feature is that, to some extent, they are within the constraints posed by the equienergy curves of the time-averaged KH potential, roughly following their shapes. Furthermore, an inspection of the classical equienergy curves shows that there is no spatial confinement beyond the separatrix, but that the momentum ranges are restricted and very small. This is specific to the atomic potential used in this work, as we have verified that for a soft-core potential with the parameters used in \cite{Norman2015}, there will be a larger momentum spread for Wigner functions coherent superpositions of KH eigenstates (not shown). Still, the momentum regions occupied by the quasiprobability flow is bounded, with vanishing net drift momentum.  This relates to the phase-space studies using classical methods \cite{Sundaram1992,Norman2015,Floriani2024}, and the computation of bounds for the quantum mechanical ionization probability \cite{Fring1996,Faria1998b}, which refer to vanishing drift momentum as a pre-requisite for stabilization.

A wave packet constructed using an equally weighted coherent superposition of these eigenstates, will exhibit a cyclic motion, which resembles that observed for molecules in our previous publications \cite{Chomet2019,Chomet2022}. The frequency associated with the motion is proportional to the energy difference between the two KH eigenstates, and is captured by the periodic oscillations in the corresponding autocorrelation function. Although the initial coherent superposition of the KH eigenstates is similar to what is obtained by combining molecular eigenstates of different parity, and also leads to similar effects (cyclic motion and population trapping) to that of a molecule, the physical mechanisms behind both are different. One should note, that for molecules in intense fields, a cyclic motion associated with the Wigner quasiprobability densities only occurs for internuclear separations much smaller than the distance between the wells of the KH potential. This is caused by a confinement effect, which, in phase space, is related to nested separatrices. For intermediate internuclear distances, these separatrices open, and enhanced ionization occurs until, eventually, for internuclear distances comparable to those in the present work, the different wells in the molecule will behave independently. This means that the quasiprobability flow between them will be strongly suppressed. 

Therefore, the analogy between the KH atom and a molecule is limited in scope. For molecules, there is tunnel ionization from one center to the other and spatial confinement, while, for the KH atom, the energy ranges associated with the wave packet and with the KH eigenstates are above that of the separatrix, suggesting over-the-barrier motion. There is a lack of spatial confinement by a potential well, but the momentum remains bounded, so that a cyclic motion is observed. Specifically for the potential studied here, the nearly vanishing drift momentum leads to a localized and cyclic wave packet in the stabilization regime. In the phase space, this is related to the equienergy curves not closing for the KH potential, but, instead, representing a boundary for the motion, which remains within the region of small momenta. Other atomic potentials may lead to larger instantaneous momenta and a more convoluted motion for the quasiprobability flow, but the momentum remains bounded if stabilization has been attained. Recently, the lack of spatial confinement in the KH atom being a limitation for its analogy with a molecule has been highlighted in the context of pump-probe schemes \cite{Ivanov2022}.

Once the full time-dependent system is taken into consideration, there is still a cyclic motion, whose periodicity, however, changes with regard to that observed for the time-averaged potential. The oscillations are well characterized, both in phase space and using the autocorrelation function. In addition to that, the Wigner functions exhibit tails associated with ionization. There has been some debate regarding their physical origin and it has been attributed to scarring \cite{Watson1995}. However, the negative quasi-probability flows and the fringes in the Wigner function indicate that the quantum interference of time-delayed ionization processes plays a role (see, for instance, \cite{Czirjak2000,Zagoya2014}. Interestingly, for all such tails, there is an increase in the electron momentum, and the Wigner quasiprobability distributions are no longer confined by the equienergy curves. This supports the view that vanishing momentum transfer is important for stabilization \cite{Fring1996,Faria1998c,Norman2015,Floriani2024}. Once more, our investigations indicate that ionization can be minimized by preparing the system in the ground state of the KH potential. 

Finally, an interesting question would be whether the level of suppression of ionization and spatial localization that we have observed is good enough for building a theoretical model of a qubit. The feasibility of this approach will depend on the robustness of this localization, precise laser field control and how well the electron's motion can be kept in bounded phase-space regions and controlled in a scalable way without leading to excessive decoherence. 
This will form the basis of future investigations. 

\noindent\textbf{Acknowledgements:} This research has been funded by the UK Engineering and Physical Sciences Research Council (EPSRC) (grant No.EP/T019530/1), by UCL and by the Institute of Physics (IoP) Bell Burnell fund.

\end{document}